\documentclass[aps,pre,amsmath,amsfonts,amssymb,floatfix,superscriptaddress,nofootinbib]{revtex4}
\usepackage{mathrsfs} \usepackage{graphicx,color} \usepackage{bbm} \usepackage{multirow}

\usepackage{rotate}
\usepackage{epsfig}
\usepackage{psfrag}
\usepackage{mathtools}

\newcommand\be{\begin{equation}}
\newcommand\bea{\begin{eqnarray} \nonumber }
\newcommand\ee{\end{equation}}
\newcommand\eea{\end{eqnarray}}

\begin{document}

\title{On the emergence of an ``intention field'' for socially cohesive agents}

\author{Jean-Philippe Bouchaud}
\affiliation{CFM, 23 rue de l'Universit\'e, 75007 Paris, France, and Ecole Polytechnique, 91120 Palaiseau, France.}
\author{Christian Borghesi}
\affiliation{Laboratoire de Physique Th\'eorique et Mod\'elisation, UMR-8089 CNRS-Universit\'e Cergy Pontoise, France.}
\author{Pablo Jensen}
\affiliation{Laboratoire de Physique, UMR-5672 CNRS, ENS de Lyon, France and Institut Rh\^onalpin des Syst\`emes Complexes, IXXI, 5 Rue du Vercors, F-69007 Lyon, France.}

\begin{abstract} 
We argue that when a social convergence mechanism exists and is strong enough, one should expect the emergence of a well defined ``field'', i.e. a slowly evolving, local quantity around which individual attributes fluctuate in a finite range. This condensation phenomenon is well illustrated by the Deffuant-Weisbuch opinion model for which we provide a natural
extension to allow for spatial heterogeneities. We show analytically and numerically that the resulting dynamics of the emergent field is a noisy diffusion equation that has a slow dynamics. This random diffusion equation reproduces the long-ranged, logarithmic decrease of the correlation of spatial voting patterns empirically found in \cite{Borghesi,Borghesi2}. Interestingly enough, we find that when the social cohesion mechanism becomes too weak, cultural cohesion breaks down completely, in the sense that the distribution of intentions/opinions becomes infinitely broad. No emerging field exists in this case. All these analytical findings are confirmed by numerical simulations of an agent-based model.
\end{abstract} 

\maketitle

\section{Introduction} 

The understanding that unremarkable individual elements can give rise, through interactions, to remarkable collective emergent phenomena is arguably  
the most important contribution of statistical physics to science. These collective effects are often so ordinary that we do not think of them as 
surprising, like for example the rigidity of a solid, which is still one of the most remarkable properties of interacting atoms \cite{PWA}. 
Some exotic states of matter are more astonishing, such as superfluidity or superconductivity, liquid crystals, etc. But this concept outreaches 
far beyond physics and is relevant to understand a host of situations, ranging from avalanches \cite{Sethna}, the functioning of the brain \cite{brain}, traffic jams \cite{Helbing}, bird flocks \cite{Cavagna}, to all kinds of social phenomena, like crowd motions \cite{Helbing-crowd}, fads \cite{Michard}, bubbles, panic and crashes, see e.g. \cite{Shelling,Buchanan,Ball,Fortunato,Crises} for recent reviews.

Recently, two of us (CB \& JPB) analyzed the spatial correlations of the turnout rate in many elections and in different European countries \cite{Borghesi,Borghesi2}. 
We found in all cases a slow, {\it logarithmic decay} of the correlation function as a function of distance. This logarithmic behaviour was very recently confirmed 
on US data by another group \cite{spain-new}. This empirical regularity not only means that the behavioral pattern of individuals is far from random, 
but also suggests that some universal underlying mechanism must be at play for such a non-trivial characteristic pattern to appear.

Intrigued by this logarithmic correlation function, we proposed in \cite{Borghesi} an interpretation in terms of a space dependent collective field, that we called a
``cultural'' field, that determines the local propensity of individuals to vote. We argued that this field ``transcend individuals while being shaped, shared, transported
and transformed by them'', and conjectured \cite{Borghesi} that the spatio-temporal evolution of this field is governed by a noisy diffusion equation 
(aka the Edwards-Wilkinson equation \cite{EW})\footnote{A similar proposal was actually made in 2000 by Schweitzer and Holyst in a different context, and without the noise component \cite{Schweitzer}.}, 
that naturally leads to logarithmically decaying correlation functions. Still, this proposal was only based on analogies and plausibility, but a detailed, agent based justification
of the existence of such a field, and its diffusive character, was lacking. The arguments put forward in \cite{Borghesi} were targeting the statistical physics community but
were deemed unconvincing in other circles, in particular sociologists, who in fact strongly balked at the name ``cultural'', if not at the concept itself. 
There are two points of possible contention. First, physicists' notion of a field is quite different from its counterpart in, for example, Bourdieu's sociology. Although in both cases the field is supposed to induce some agents' actions, in Bourdieu's meaning of the word \cite{Bourdieu} the field designates a much more heterogeneous structure, a system of social positions structured by power relationships in which different actors occupy different positions and try to seize social capital (such as money or influence) according to their specific position in the ``field''. Here, the field is 
rather a scalar quantity, slowly changing in space. Second, the concept of a ``field'' as an autonomous entity automatically inducing agents' actions, as if these were passive puppets, is criticized by some sociologists \cite{Latour}.

The aim of the present work is precisely to provide a well-defined, agent based micro-foundation to the notion of field (that we rename henceforth ``intention field'', 
to avoid any possible confusion induced by the word ``cultural''). As we were working on these ideas, ref. \cite{spain-new} appeared on arXiv with  related, 
but different ideas. Here, we investigate a model of interacting agents that exhibits a transition between a socially cohesive phase, where the concept of an emergent 
intention field has a precise meaning and obeys a diffusion equation, and a socially disconnected phase where no emergent quantity can be defined. The model we consider can be viewed as a spatially heterogeneous version of the Deffuant-Weisbuch model \cite{Weisbuch}. When simulated on the actual map of France towns, our model is found to reproduce {\it quantitatively} the shape of the correlation function measured on data. Our model furthermore allows us to discuss several interesting questions, such as the role of population migrations on the integrity of the intention field. 

\section{The model}

We assume each potential voter $i$ in city $\alpha$ (located in space around $\vec R_\alpha$) has, at time $t$, a propensity to vote (or ``intention'') $\phi_{\alpha,i}(t)$. The probability that this potential voter will actually vote on election day is chosen to be the logistic function, as is standard in choice theory \cite{Anderson}:
\be
p = \frac{1}{1+e^{-\phi}}.
\ee
Now, we posit, as in many previous studies \cite{Weisbuch,spain,spain2,Chakraborti,Fortunato} that the intention $\phi$ of each individual evolves through interactions with others. (This can of course be generalized to other situations beyond vote, 
and the intentions can be also thought of as ``opinions''). 

More precisely, we postulate the following rules, that are quite general and plausible. Between $t$ and $t+dt$, each individual $i$ of city $\alpha$ can meet an individual $j$ from city $\beta$, and his intention at the end of the meeting will have evolved towards that of individual $j$, by some quantity, possibly zero. The probability (per unit time) that these two individuals meet {\it and} that the encounter leads to a change of intention is:\footnote{We assume for simplicity that all cities have
the same population size, and identify number of individuals and probabilities.}
\be
w(i,\alpha ; j, \beta)=  f(|R_\alpha - R_\beta|) \, G\left[|\phi_{\beta,j} - \phi_{\alpha,i}|\right],
\label{eaccroch}
\ee
where $f(r)$ is a decreasing function (over a scale of a few kms) and  
$G$ a decreasing function of its argument, as to model, in the spirit of the Deffuant-Weisbuch model \cite{Weisbuch}, the fact that far away 
opinions have a smaller probability to be in the same social circle, or that such encounters leave each participant with unchanged opinion. 
The quantity $dt \sum_{\beta,j} f(|R_\alpha - R_\beta|) G[|\phi_{\beta,j} - \phi_{\alpha,i}|]$ 
is the total probability of an encounter for individual $i$ between $t$ and $t+dt$. In practice, a reasonable order of magnitude is probably one (or a few) per week.

The result of this encounter is that both intentions will evolve as:\footnote{Many variations are of course possible, such as choosing $\gamma$ to be a random variable, but we believe that our results are robust to most changes. Note in particular that the rule given by Eq. (\ref{eop-dyn}) does not conserve the total intention at
each encounter.}
\bea
\phi_{\alpha,i} &\to& \phi_{\alpha,i} + \gamma (\phi_{\beta,j} - \phi_{\alpha,i}) + \eta \\
\phi_{\beta,j} &\to& \phi_{\beta,j} + \gamma (\phi_{\alpha,i} - \phi_{\beta,j}) + \eta' ,
\label{eop-dyn}
\eea
where $0 \leq \gamma \leq \frac12$ measures how intentions get closer together, much like in other opinion formation models, 
and $\eta,\eta'$ describe all sources of noise due to other influences, that vary both in time and across agents, with some distribution $Q(\eta)$, for example Gaussian. 
The component of $\eta$'s common to all agents, independently of the city where they live, can be reabsorbed in a shift of all $\phi_{\alpha,i}$ by 
a time dependent contribution $\overline{\phi}(t)$, which is irrelevant in the following discussion. We will assume henceforth that the remaining idiosyncratic $\eta$'s 
are independent, identically distributed variables, both in time and across agents. (In a refined version of the model, it might be interesting to allow $\eta$ 
to be correlated for individuals of the same city, in particular to account for the empirical dependence of the variance of $\phi$ on the city size, reported in \cite{Borghesi2}.)

These rules allow one to write a Boltzmann-like equation for the probability $P_t(\phi|\alpha)$ that a given individual in city $\alpha$ has intention $\phi$:
\be\label{Boltzeq}
\frac{\partial P_{t}(\phi|\alpha)}{\partial t} = - W_t(\phi) P_{t}(\phi|\alpha) + \sum_\beta f(|R_\alpha - R_\beta|) \iiint d\eta d\psi d\psi' 
Q(\eta) P_{t}(\psi|\alpha) P_{t}(\psi'|\beta) G(|\psi-\psi'|)\delta\left(\phi - \psi - \gamma (\psi' - \psi) - \eta \right),
\ee 
with
\be
W_t(\phi) = \sum_\beta f(|R_\alpha - R_\beta|) \int d\psi' P_{t}(\psi'|\beta) G(|\phi-\psi'|).
\ee
By inspection, one checks that the above Boltzmann equation preserves the total probability in each city (at this point 
we do not consider the possibility that people may change places -- but see below). 

In the spirit of the Chapman-Enskog expansion that allows one to derive (macroscopic) Navier-Stokes equations from the (microscopic)
Boltzmann equation \cite{Boltzmann}, we will argue that under some conditions, a local equilibrium is quickly reached, of the form:
\be
P_{eq}(\phi|\alpha) = P^*(\phi-\varphi_\alpha),
\ee
where $\varphi_\alpha = \int d\phi \phi P_{eq}(\phi|\alpha)$ is the (time dependent) average of the intention of the individuals in city $\alpha$, for which a diffusion equation can be established.
The emergence of a slow ``intention'' field $\varphi_\alpha$ (which is what we called the cultural field in \cite{Borghesi}) is the analogue of the 
emergence of the macroscopic velocity field $\vec U(\vec r,t)$ in the Boltzmann equation. In this case, the distribution of velocities quickly reaches
a local equilibrium shape (the Maxwell distribution) centered around $\vec U(\vec r,t)$ that obeys a (slow) hydrodynamic equation. 
As with the intention/opinion field, each molecule in the hydrodynamic flow interacts with the velocity field $\vec U(\vec r,t)$ such that the local equilibrium is
maintained, while the latter is an emergent, collective variable (the local velocity average) that does not ``belong'' to any molecule in particular,
but acquires an autonomous, slow dynamics that allows it to persist on long time scales and develop long range correlations (whereas the residual velocity of each molecule/individual 
decorrelates on fast time scales).

\section{Local equilibrium and phase transitions} 

Let us first look for a homogeneous solution of the Boltzmann equation, with $\varphi_\alpha=\varphi^*$ that can, at this point, be set to zero without loss of generality:
\be
P_{eq}(\phi|\alpha) = P^*(\phi).
\ee
In order to keep the calculations tractable, we assume that $\eta$ are Gaussian variables of zero mean and variance $\Sigma^2$, and that the 
interaction kernel $G$ is also Gaussian with a range $\zeta$:
\be
G(u) = e^{-u^2/2\zeta^2},
\label{eGu}
\ee
meaning that individuals with intentions/opinions that are too distant from each other ($> \zeta$) cease to interact, 
much as in the original model of Deffuant \& Weisbuch \cite{Weisbuch}. We believe that these specific choices do not change the qualitative conclusions below.

We will also write:
\be
 \sum_\beta f(|R_\alpha - R_\beta|) = F
\ee
that could depend on $\alpha$ for non translation invariant geographies, but this would only change the local equilibration time and not the shape of the equilibrium. Note that $F$ has the dimension of an inverse time scale.

We look for a Gaussian solution:
\be
P^*(\phi) = \frac{1}{\sqrt{2\pi \sigma^2}} e^{-\phi^2/2\sigma^2}.
\ee
Inserting into the definition of $W(\phi)$ yields:
\be
W^*(\phi) = F \frac{\zeta}{\sqrt{\zeta^2+\sigma^2}} e^{-\frac{\phi^2}{2(\zeta^2+\sigma^2)}}.
\ee
Hence, $W^*(\phi)P^*(\phi)$ is proportional to a Gaussian of zero mean and variance $\sigma^2(\zeta^2+\sigma^2)/(\zeta^2+2\sigma^2)$.

Now, look at the second term in the right-hand side of the equation:
\be
\iiint  d\eta d\psi d\psi' 
Q(\eta) P^*(\psi) P^*(\psi') G(|\psi-\psi'|)\delta\left(\phi - \psi - \gamma (\psi' - \psi) - \eta \right),
\ee
which is clearly also a Gaussian in $\phi$ with zero mean and variance:
\be
\Sigma^2 + \frac{\sigma^2}{1+2\Gamma}(1+\Gamma-2\gamma(1-\gamma)), \qquad \Gamma = \frac{\sigma^2}{\zeta^2}.
\ee
Equating this expression with the above result leads to an equation for our unknown $\sigma^2$:
\be\label{transition-eq}
\sigma^2 = \frac{1}{2} \frac{\Sigma^2 \zeta^2}{\gamma(1-\gamma)\zeta^2 -\Sigma^2},
\ee
which, obviously, only makes sense if the denominator is positive. Therefore, a stationary (Gaussian) solution only exists provided the idiosyncratic noise is small enough:
\be
\Sigma^2 < \Sigma^2_c=\zeta^2 \gamma (1-\gamma).
\ee
We show in the Appendix that for larger idiosyncratic noise, there is indeed {\it no} stationary state to the Boltzmann equation: the variance of $P(\phi)$ diverges without bound (in the 
absence of any other mechanism limiting the growth of $\phi$) and there is therefore no emergent average opinion/intention. 

Below this value, on the other hand, the opinion-sharing interaction between individuals is strong enough to keep the dispersion of intentions finite. The above inequality means 
that for this to happen the idiosyncratic evolution of intentions (of strength $\Sigma$) must be small compared to the convergence effects: larger $\gamma$ 
leads to a stronger reduction of heterogeneities at each encounters, while $\zeta$ large means that even far away opinions have a chance to meet and converge. Our 
numerical results below fully confirm these analytic findings.

The above transition between confined and dispersed intentions, i.e. between consensus and dissent, is in fact generic. 
It has been noted in the context of the Deffuant-Weisbuch model in \cite{spain} (see also \cite{spain2,Chakraborti}).
It is also present in a variation of the above model, where intentions evolve not only after an encounter, but further diffuse continuously in time, due to influences unrelated to encounters, like
all sorts of books, news media, personal observation and experiences, etc. which contain a part common to all agents (which can be discarded as noted above) and a part that is agent specific. This 
last contribution adds a diffusion term in the Boltzmann equation of the form $D \partial^2 P(\phi)/\partial \phi^2$. We again find a phase transition in this case, although it is now a {\it first order transition}, where $\sigma^2$ jumps discontinuously from a finite value when the diffusion constant $D$ is smaller than
a certain threshold $D_c \propto \gamma(1-\gamma)F\zeta^2$, to infinity for larger idiosyncratic noises. The transition discussed above, on the other hand, is second order, with 
$\sigma^2$ diverging continuously as $\Sigma^2$ reaches the threshold value $\Sigma^2_c$, see Eq. (\ref{transition-eq}). 

In any case, the conclusion of this analysis is that when the idiosyncratic noise (called -- perhaps ironically? -- ``free will'' in \cite{spain}) is small enough, the basic 
message of the Deffuant-Weisbuch model holds: there is convergence of the propensities, intentions or opinions of the different individuals towards the average initial opinion of the population, 
up to a random variable of zero mean (for infinite size populations) and variance given by $\sigma^2$.  The  
{\it condensation of opinions} emphasized by Weisbuch et al. is essentially what is needed for a collective variable (the average $\varphi_\alpha$) to emerge and play a special role in the long term evolution of the system. In this sense, our model is a spatially heterogeneous extension of the noisy Deffuant-Weisbuch model. Beyond the threshold, there is no convergence anymore and the concept of an intention field (the average intention or opinion) evaporates and becomes meaningless. This happens when the social interactions are no longer strong enough to maintain cultural cohesion.

The convergence of $P(\phi)$ towards its equilibrium, condensed shape is however non trivial and needs to be discussed. When the width of initial distribution $\sigma_0$ is not too broad compared
to the interaction distance $\zeta$, the time $\tau$ needed for the initial distribution to converge towards its asymptotic form $P^*(\phi)$ is set by a {\it renormalized} microscopic time scale, given by:
\be
\tau^{-1} = F \frac{\zeta}{\sqrt{\zeta^2+\sigma^2}}.
\ee
In the consensus phase far from the transition $\Sigma^2 \ll \zeta^2 \gamma (1-\gamma)$, one has also $\sigma^2 \ll \zeta^2$, and therefore $\tau \approx F^{-1}$, where
$F$ is simply the total ``encounter rate'' for each individual, which is, as we said above, roughly one per week. Therefore, we expect convergence to happen quickly, 
over weeks or perhaps months in a well connected society, where individuals share their thoughts. If on the other hand the system approaches its deconfining phase transition, $\sigma^2 \gg \zeta^2$
and the equilibration time diverges as $\tau \approx F^{-1} \sigma/\zeta$. 

The situation becomes quite interesting when the initial condition becomes broad, i.e. $\sigma_0 \gg \zeta$. In this case, the convergence takes place, in the condensed phase, in two steps. First, after a time of order $F^{-1}$, several sharp peaks emerge, corresponding to a mixed phase where sub-populations are socially cohesive, but the population as a whole is fragmented \cite{Weisbuch}. Then, 
a very slow process sets in, through which the number of these peaks progressively decreases and the width of $P(\phi)$ finally decreases to its equilibrium value $\sigma^2$. However, the time 
needed for this to happen can be estimated to be $\sim e^{\sigma_0^2/\zeta^2} \ln N$, which can be so long that equilibrium concepts become irrelevant, and several ``sub-cultures'' survive in the
population. This picture is fully confirmed by our numerical results, see Fig.~\ref{fCI} below. 

\section{Spatial heterogeneities}

At this point, we borrow from the standard Chapman-Enskog approach to the Boltzmann equation \cite{Boltzmann} and assume that a) local equilibration towards a space dependent, shifted 
distribution $P^*(\phi - \varphi_\alpha)$ is fast and b) the field $\varphi_\alpha$ varies smoothly and slowly in space, in such a way that 
$\varphi_\alpha - \varphi_\beta$ is small whenever two individuals from the two cities have an appreciable probability to interact (i.e. when $f(|R_\alpha - R_\beta|)$ is large).

The time evolution of $\varphi_\alpha$ can then be computed from the Boltzmann equation, only retaining contributions linear in $\varphi_\alpha - \varphi_\beta$ and discarding terms of higher order in the difference. The final results reads:
\be\label{diff-eq}
\frac{\partial \varphi_\alpha}{\partial t} =  \frac{\zeta}{\sqrt{\zeta^2+2\sigma^2}}\left(1 + (\gamma -\frac12)\frac{\zeta^2}{\zeta^2+2\sigma^2}\right)
\sum_{\beta} f(|R_\alpha - R_\beta|) (\varphi_\beta-\varphi_\alpha), 
\ee
which is precisely the diffusion equation proposed in \cite{Borghesi,spain-new}. This allows one to obtain the value of the diffusion constant in terms of the 
microscopic parameters of the model. If the range of the function $f$ is $\xi$, we find, in order of magnitude, that the diffusion constant is 
given by $\xi^2/\tau$, where $\tau$ is the equilibration time discussed above. The relaxation time of the $\varphi$ field, in a country of linear size $L$, is
therefore $\sim \tau (L/\xi)^2$ which is much larger than the local equilibration time itself provided $L \gg \xi$. For France, for example, $L \sim 500$ km while 
$\xi$ is maybe $5$ km or so, so that the ratio of time scales is $\sim 10,000$! It is precisely this separation of time scales that justifies the very idea 
of a ``hydrodynamical'' description in terms of a collective field. 

In the above derivation, we have not taken into account the noise term that describes city-specific contributions to the intentions. This would add an additive random term
to the above equation, which, as shown in \cite{Borghesi}, eventually leads to the logarithmic spatial correlation profile found empirically. 

The microscopic model considered up to now discards several effects that would be quite easy to reintegrate. One interesting effect is immigration of 
individuals with significantly different cultures. This can be seen as adding to the Boltzmann equation a term of the form:
\be
\left.\frac{\partial P_{t}(\phi|\alpha)}{\partial t}\right|_{imm.} = \nu \left[\rho(\phi)-P_t(\phi|\alpha)\right], 
\ee
where $\nu$ is the immigration rate and $\rho(\phi)$ is the distribution of intention (/opinion) of the incoming population. If $\nu$ 
is small enough, convergence towards a local, single peak equilibrium distribution will still hold. Although the local average $\varphi_\alpha$ evolves towards
the average intention of the new population, cultural cohesion is maintained. This is similar to the
observation by Grauwin and Jensen in \cite{Grauwin}, where they show that for small enough population turnover, ``structures may last longer than 
individual entities''. This is precisely saying that there exist a slowly evolving local quantity around which individual attributes (propensity, intention, opinion, etc.) 
fluctuate in a finite range. For larger values
of $\nu$, on the other hand one may expect a phase transition akin to what happens in the original Deffuant-Weisbuch model, where a multimodal distribution of 
intentions or opinions sets in. In this case, a description in terms of a unique field is clearly not warranted. 

Another effect is that people can change places, adding a true transport term in the above Boltzmann equation, of the form:
\be\label{move-eq}
\left.\frac{\partial P_{t}(\phi|\alpha)}{\partial t}\right|_{tr.} = - \sum_{\beta} W_{\alpha \to \beta} P_t(\phi|\alpha) 
+ \sum_{\beta} W_{\beta \to \alpha} P_t(\phi|\beta) ,
\ee
where $W_{\alpha \to \beta}$ is the probability per unit time that an inhabitant of city $\alpha$ moves to city $\beta$. This extra term may lead to
two different effects. An obvious one is to increase the effective diffusion constant in equation (\ref{diff-eq}) above, or possibly even change 
the diffusion structure by allowing long-range, L\'evy-like jumps (for example 
from Paris to Montpellier, which has become popular move recently).\footnote{Here we discard the possibility that the destination of the move 
might be correlated with the initial opinion of the individual. But much as the $G$ function decays as the opinion ``distance'' increases, it might 
well be that the transition rate $W_{\alpha \to \beta}$ depends on $\phi - \varphi_\beta$.} 
However, we believe, on the basis of the numerical results reported below, that this effect is much smaller than the diffusion induced by the contact process investigated above, at least 
in European countries where internal mobility is still quite small. Indeed, whereas contact interaction takes place on the scale of weeks, changing places rather 
happens on the scale of several years. Still, this could play an stronger role in the US for example. 

The other effect of the extra term (\ref{move-eq}) is similar to the immigration
process above. If the moving process is sufficiently frequent, mixing between different 
populations maybe strong enough to impede the convergence towards any local equilibrium. In this case, the whole construction above, based on the emergence of 
a local field, should break down. This, on the other hand, is not unexpected when mixing becomes so strong that the notion of locality becomes moot. We
will investigate this case numerically in the next section.

\section{Numerical simulations}

\begin{figure}[t]
\includegraphics[width=8cm, height=6cm,clip=true]{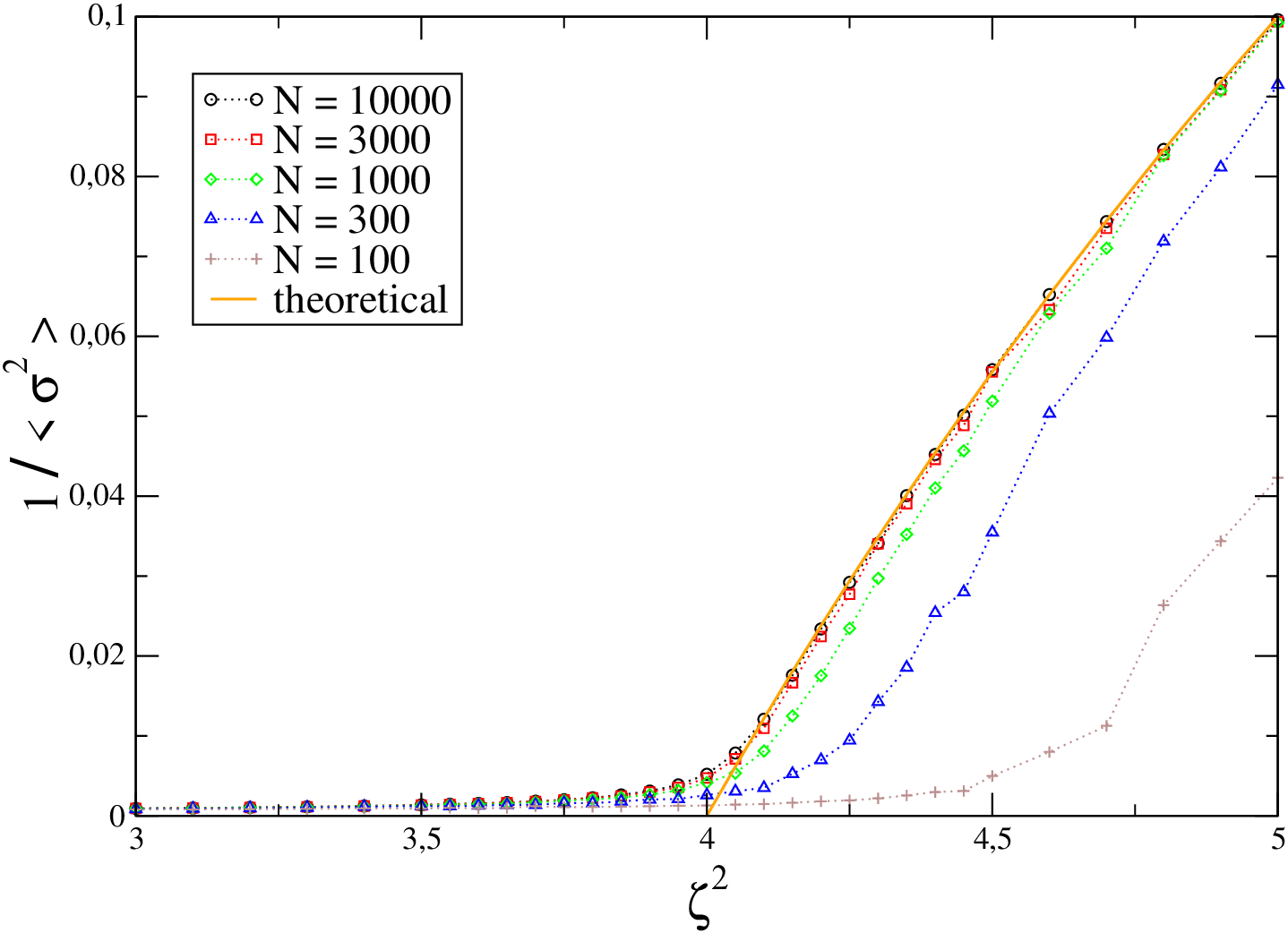}
\caption{\small  Homogeneous model, corresponding to single city dynamics. We plot ${\langle \sigma^2 \rangle}^{-1}$ as a function of $\zeta^2$ for different system-sizes $N$. The theoretical value given by Eq.~(\ref{transition-eq}), and corresponding to $N = \infty$, is also plotted. The agreement is excellent, once finite size effects are accounted for.}
\label{fintra}
\end{figure}

We now present numerical simulations of our model, with three aims in mind. First, we want to confirm the existence of the second order condensation transition
that we obtained analytically above, and study the equilibration dynamics. Second, we want to investigate the spatial pattern generated by our model when simulated with the real positions of French 
towns, and a short range interaction kernel $f(r)$, and demonstrate that the logarithmic behaviour of the correlation function indeed appears. Finally, we 
investigate the role of moving places, and show that beyond a certain frequency of long range moves, the whole correlation pattern vanishes.

\subsection{Single city dynamics}

We first consider the case where all $N$ agents live in the same city. For each iteration, a pair of individuals $i,j$ is chosen at random and change their opinion/intention $\phi_{i}$ and $\phi_{j}$ according to Eq.~(\ref{eop-dyn}) with the probability given by Eq.~(\ref{eaccroch}), with $f(r=0)=1$. We choose 
$\gamma=0.5$ and fix the idiosyncratic noise variance $\Sigma^2=1$. We study the system as a function of the width of the interaction kernel $\zeta$, see Eq.~(\ref{eGu}). 
Note that the critical value for $\gamma=0.5$ and $\Sigma^2=1$ is $\zeta^2_c=4$.

From the calculations above, we predict a transition towards a condensed phase where the variance $\sigma^2$ of the opinion/intention becomes finite when $\zeta^2 > \zeta_c^2=4$. Figure~\ref{fintra} shows ${\langle \sigma^2 \rangle}^{-1}$ as a function of $\zeta^2$ for different system sizes $N$, and with initial condition $\sigma_0^2=1$. The average of $\sigma^2$ is made over 50 realisations long after the transients. Theses curves are compared with the theoretical prediction, Eq.~(\ref{transition-eq}), with very good 
agreement when $N$ is sufficiently large. In particular, our numerical results are in full agreement with the existence of a second order  
transition for $\zeta^2_c=4$, beyond which a socially cohesive situation sets in. 

\begin{figure}[t]
\includegraphics[width=8cm, height=6cm,clip=true]{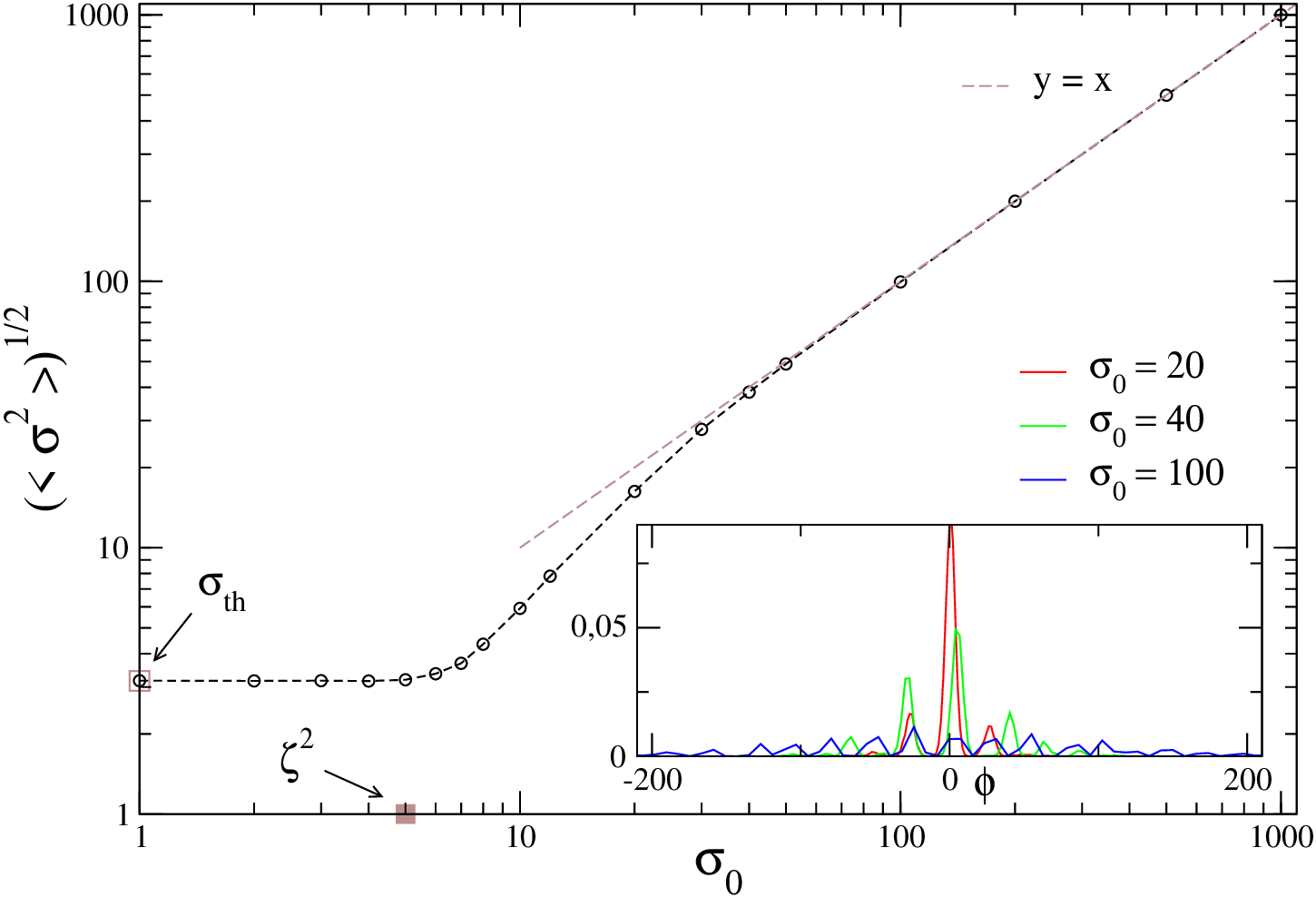}
\caption{\small Role of the initial conditions. We now plot ${\langle \sigma^2 \rangle}^{1/2}$ as a function of the initial width of the distribution $\sigma_0$, measured after a time 
$t=40000$. For $\sigma_0 <\sim \zeta$, the result agrees with the equilibrium theoretical prediction, Eq.~(\ref{transition-eq}), as expected. But for $\sigma_0 \gg \zeta$, the distribution fragments 
into sub-peaks (as shown in the inset), but the overall width of the distribution hardly shrinks with time. The time needed for convergence becomes exceedingly long in this case, and 
equilibrium may lose its relevance.}
\label{fCI}
\end{figure}

When the initial condition becomes broader, however, we observe the fragmentation of the population discussed above and first reported in \cite{Weisbuch}. The width of the distribution $P(\phi)$ 
obtained after a long, but finite time $t=40000$, is shown in Fig.~\ref{fCI}. 

\subsection{Spatial correlations}

Spatial heterogeneities are taken into account via the real positions of the {\it mairie} (town-hall) of the $36,000$ French mainland municipalities. We consider that all agents living in a given municipality have exactly the same spatial position: that of the {\it mairie}~\cite{Borghesi,Borghesi2}. Moreover, in order to test spatial effects we assume that each municipality has the same number of agents, $N=1000$. This is quite small in view of Fig.~\ref{fintra}, but this already
corresponds to a simulation with $36$ million agents, and we found little differences with the case $N=100$. However, when computing the total variance of $\phi$'s over all cities, we already find a result very close to the theoretical 
prediction, Eq.~(\ref{transition-eq}). This comes from the fact that agents of nearby cities interact, leading to a large effective value of $N$. In any case, our main focus here is to demonstrate the emergence of a logarithmic dependence of the spatial correlation function.

\begin{figure}[t]
\includegraphics[width=8cm, height=6cm,clip=true]{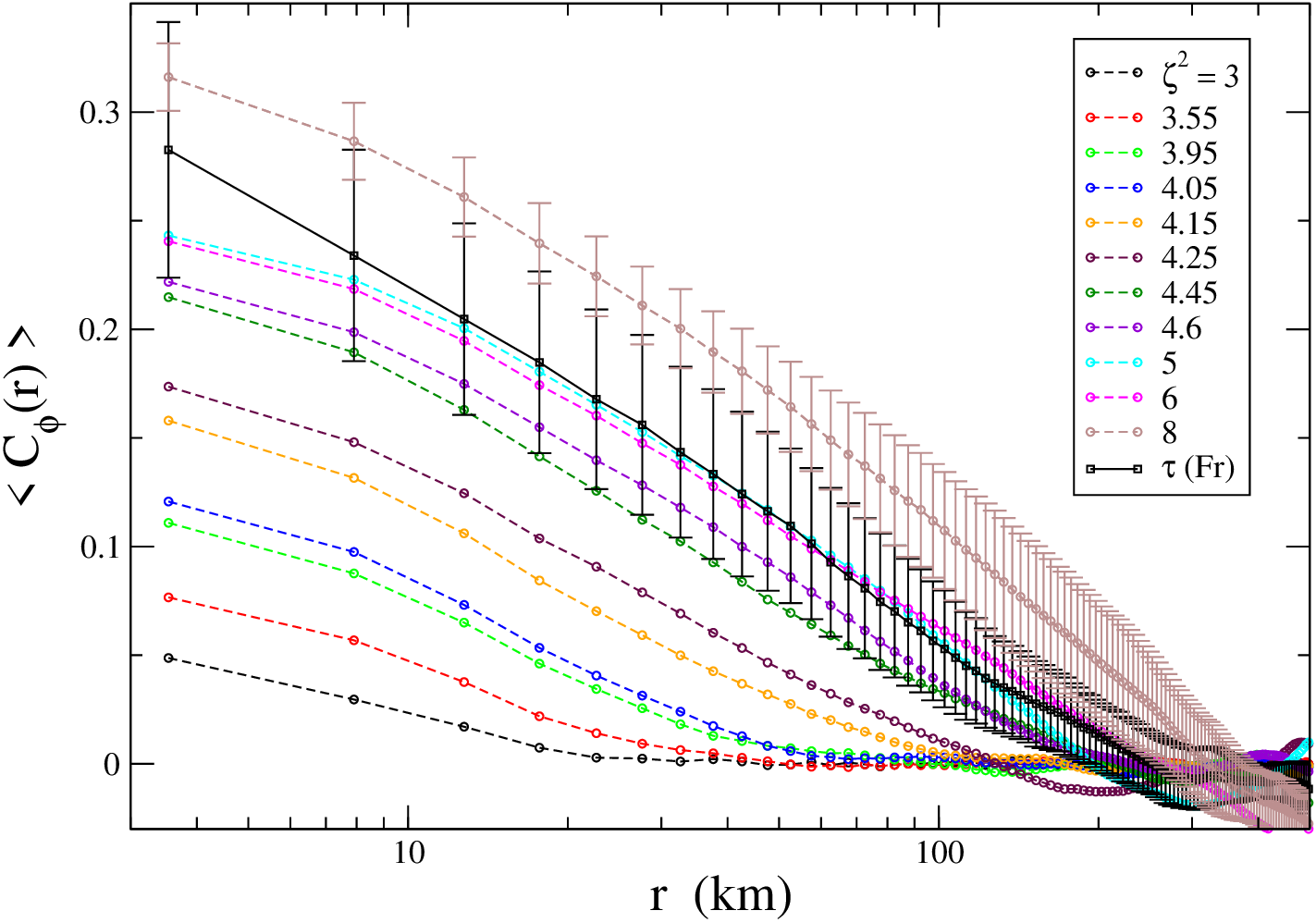}
\caption{\small Average spatial correlations $\langle C_\phi(r) \rangle$ of $\phi$ as a function of $r$ for different values of $\zeta^2$, with $N=1000$ agents in each city. We show for comparison 
the average spatial correlation of the empirical logarithmic turnout rate, for the same $36,000$ municipalities over 20 French national elections. The logarithmic behaviour 
of the empirical correlation function is very well accounted for with our agent based model into its socially cohesive phase: $\zeta^2 \approx 5 - 6$.}
\label{fspatial}
\end{figure}

We choose the spatial interaction kernel $f(|R_\alpha - R_\beta|)$ as an exponential, $f(r) = \exp(-r/\xi)$. The characteristic length $\xi$ is chosen to
be the same as in \cite{Borghesi,Borghesi2}: $\xi=4.5\,\mathrm{km}$.  Fig.~\ref{fspatial} shows, for different values of $\zeta^2$, the average spatial correlations, defined exactly as in \cite{Borghesi,Borghesi2} for 
the empirical data:
\be
C_\phi(r)  = \frac{\left.\left\langle (\phi_\alpha - m)(\phi_\beta -m) \right\rangle\right|_{r_{\alpha\beta}=r}}{\langle (\phi_\alpha-m)^2 \rangle},
\ee
where the average is taken over all cities $\alpha,\beta$ at a certain distance $r$ (within a certain $dr=\pm 2.5\:\mathrm{km}$) and $m= \langle \phi_\alpha \rangle$ is the (national) average value of $\phi$. 
The correlation function is furthermore averaged over 50 realisations, after equilibration
is reached. We also show the average spatial correlation of the actual logarithmic turnout rate, for the same $36,000$ municipalities over the 20 French national elections 
studied in~\cite{Borghesi2}. As expected, the overall level of $C_\phi(r)$ increases with $\zeta^2$. More interestingly, we see a clear logarithmic
dependence of $C_\phi(r)$ as a function of $r$, that extrapolates to zero for $r \sim 300$km, exactly as for the empirical results. We see that when
the model is well into the cohesive phase ($\zeta^2 \approx 5 - 6$), the value and shape of $C_\phi(r)$ is well reproduced, without any extra fitting 
parameter. The main message is that our spatial extension of the Deffuant-Weisbuch model is indeed able to account for the logarithmic decay of spatial correlations. This is in line with the conclusions of \cite{spain-new} as well.
 
\subsection{The influence of changing places}

Up to now, we have considered exchange of opinion or intention between agents living relatively close-by, since we assume that the interaction kernel $f(r)$ 
decays on the scale of a few kilometers. This is probably reasonable for most, every-day life encounters. Still, this model neglects vacations or business travels, 
where people can meet with family, friends or colleagues who live very far apart. This could be modeled by adding a slowly decaying tail to $f(r)$. Another effect 
is that people occasionally change places; these moves can be long-ranged and are permanent. Intuitively, it is clear that the effect of these moves is to
strongly mix the opinions/intentions; this could destroy the logarithmically correlated spatial patterns discussed above. 

In order to investigate this effect, we have added the following ingredient to our numerical simulation. At each time step, each agent is given a certain 
probability $\pi$ to move. If he/she decides to move, the new town is chosen randomly among the $36,000$ municipalities. This means that most moves in this model are long-ranged, since the average distance of a move is on the order of $L$, the size of the country. The move is in fact a swap, such as the number of inhabitants per city is fixed throughout the simulation. The correlation function $C_\phi(r)$ is then measured for 
different values of $\pi$. We find is that whenever $\pi < \pi_c$, the correlation remains approximately logarithmic, but with an amplitude and a range 
that become smaller and smaller:
\be
C_\phi(r) \approx A(\pi) \left[\ln L^*(\pi) - \ln r\right], \quad r < L^*(\pi); \qquad C_\phi(r) \approx 0, \quad r > L^*(\pi)
\ee
where both $L^*(\pi)$ and $A(\pi)\ln L^*(\pi)$ decrease when $\pi$ increases, and appear to vanish continuously and simultaneously for $\pi = \pi_c$. 
Beyond $\pi_c$, the system is completely homogeneized and $C_\phi(r >0)$ is zero. We however find that the value of $\pi_c$ is quite large: it corresponds to
$\approx 0.03$ per agent and per time step, which, as we argued, should correspond to a week in physical time. Actual rates of moving, especially long-range 
moves, are much smaller than this, probably once every 20 years (but a fraction of the population actually never moves, so assuming a unique $\pi$ for all agents is not a good approximation). This would correspond to $\pi = 10^{-3}$, a value much smaller than the homogeneization threshold $\pi_c$. It is therefore reasonable to neglect these moves in order to account for the spatial pattern of election turnouts.

\section{Conclusion}

In conclusion, we have argued  that if an opinion/intention convergence mechanism exists and is strong enough, one should expect the emergence of a slowly evolving, local quantity around which individual attributes fluctuate in a finite range. We identify this quantity with the ``cultural field'' proposed  in \cite{Borghesi}. The opinion condensation mechanism is well illustrated by the Deffuant-Weisbuch model; our mathematical analysis builds upon this model and allows for spatial heterogeneities. We show analytically and numerically that the resulting dynamics of the emergent field is a noisy diffusion equation that has a slow dynamics, much slower than the time needed to reach local equilibrium, as required for a hydrodynamical description 
in terms of collective variables. This random diffusion equation reproduces the long-ranged, logarithmic decrease of the correlation of spatial voting patterns empirically found in \cite{Borghesi}. Interestingly enough, we find, again theoretically and numerically, that when social cohesion mechanisms are 
too weak (i.e. when individuals do not communicate enough and/or are not influenced by the opinion of others), cultural cohesion can break down completely, 
in the sense that no local equilibrium exists and the distribution of intentions/opinions becomes infinitely broad. In this case, obviously, the notion of ``field'' becomes
meaningless. Reversing the argument, though, we view the empirical results of \cite{Borghesi,Borghesi2,spain-new} on vote patterns as a strong quantitative evidence that the western populations studied there are in a culturally cohesive phase, at least as far as voting habits are concerned. 

The ``neighbourhood effect'' on voting patterns has been conjectured in political geography 
papers as early as 1969 \cite{Cox}, and substantiated by some quantitative studies that suggest ``conversion by conversation'' \cite{McAlister}. A fuller understanding of voting patterns needs to account for the strong dependence of voting attitudes on people intrinsic characteristics such as social status or age (see for example  \cite{andersen}). Since social status is known to be spatially correlated as well  \cite{jargowsky}, further studies are needed to disentangle the respective effects of social status and ``conversion by conversation'' in the spatial correlations observed. Extension of these ideas to other type of behavioral patterns would also be highly interesting.

\section*{Acknowledgements}

This work sprung out of a one-day meeting between physicists and sociologists at Sciences-Po, Paris, on the May the 24th, 2013, entitled 
``Le tout est-il plus ou moins que la somme de ses parties ?''. We thank Bruno Latour and Juliette Stehl\'e for useful comments and suggestions.

\section*{Appendix}

We establish here an equation for the dynamical evolution of the variance of the intentions $\phi$, valid for a Gaussian interaction kernel $G(u)$. We start from the Boltzmann equation, (\ref{Boltzeq}), and multiply 
both sides by $\phi$ and integrate over $\phi$ to get (assuming $\langle \phi \rangle=0$:
\be
\frac{\partial \langle \phi^2 \rangle_t}{\partial t} = F \Sigma^2 \iint d\psi d\psi' P_{t}(\psi) P_{t}(\psi') G(|\psi-\psi'|) + F \gamma(1 - \gamma) \iint d\psi d\psi' (\psi-\psi')^2 P_{t}(\psi) P_{t}(\psi') G(|\psi-\psi'|).
\ee 
Going to Fourier space with the corresponding $\widehat P_t(k)$ and $\widehat G(k)$, one has:
\be
\iint d\psi d\psi' (\psi-\psi')^2 P_{t}(\psi) P_{t}(\psi') G(|\psi-\psi'|) \equiv - \int \frac{dk}{2 \pi}  \widehat P_t(k)\widehat P_t(-k) \frac{\partial^2 \widehat G(k)}{\partial k^2}.
\ee
Using the fact that $G(u)$ is Gaussian allows one to get:
\be
- \int \frac{dk}{2 \pi}  \widehat P_t(k)\widehat P_t(-k) \frac{\partial^2 \widehat G(k)}{\partial k^2} = \zeta^2 \int \frac{dk}{2 \pi}  \widehat P_t(k)\widehat P_t(-k) \widehat G(k),
\ee
where the second term is exactly $\zeta^2 \Phi(t)$, where $\Phi(t)$ is the rate of efficient encounters per unit time:
\be
\Phi(t) = \int \frac{dk}{2 \pi}  \widehat P_t(k)\widehat P_t(-k) \widehat G(k) \equiv \iint d\psi d\psi' P_{t}(\psi) P_{t}(\psi') G(|\psi-\psi'|).
\ee
Therefore, finally:
\be
\frac{\partial \langle \phi^2 \rangle_t}{\partial t} = F \Phi(t) \left[ \Sigma^2 - \gamma(1 - \gamma) \zeta^2 \right],
\ee 
which shows that whenever $\Sigma^2 > \gamma(1 - \gamma) \zeta^2$, the variance of $\phi$ grows without bound and no stationnary state can be reached.


\begin{thebibliography}{20}

\bibitem{Borghesi} C. Borghesi and J.-P. Bouchaud, {\it Spatial correlations in vote statistics: a diffusive field model for decision-making}, Eur. Phys. J. B 75, 395-404 (2010)

\bibitem{Borghesi2} C. Borghesi, J.-C. Raynal and J.-P. Bouchaud, {\it Election turnout statistics in many countries: similarities, differences, and a diffusive field model for decision-making} PLoS ONE 7, e36289 (2012).

\bibitem{PWA} On this point, please read the deeply insightful remarks of P. W. Anderson in Basic Concepts in Condensed matter Physics Reading: Addison-Wesley (1997). One particularly relevant quote in the present context is: {\it We are so accustomed to [this] rigidity property that we don't accept its almost miraculous nature, that is an ``emergent property'' not contained in the simple laws of physics, although it is a consequence of them.}

\bibitem{Sethna} J. Sethna, K. Dahmen, C. Myers, {\it Crackling Noise}, Nature, {\bf 410}, 242 (2001)


\bibitem{brain} see e.g. L. de Arcangelis, C. Perrone-Capano, H. J. Herrmann, {\it Self-organized criticality model for brain plasticity}
Physical Review Letters 96 (2), 028107 (2005); 
E. Tagliazucchi, D. Chialvo, {\it The collective brain}, Decision Making: pp. 57-80 (2011), 


\bibitem{Helbing} D. Helbing, {\it Traffic and related self-driven many-particle systems}, Reviews of Modern Physics 73 (4), 1067
(2001)

\bibitem{Cavagna} W. Bialek, A. Cavagna, I. Giardina, T. Mora, E. Silvestri, M. Viale, A. M. Walczak, {\it Statistical mechanics for natural flocks of birds},
Proceedings of the National Academy of Sciences 109 (13), 4786-4791 (2012)

\bibitem{Helbing-crowd} D. Helbing, A. Johansson, H. Z. Al-Abideen, {\it Dynamics of crowd disasters: An empirical study}
Physical Review E 75, 046109 (2007)

\bibitem{Michard} Q. Michard and J.-P. Bouchaud, {\it Theory of collective opinion shifts: from smooth trends
to abrupt swings}, European Physical Journal B, 47:151, (2005); C. Borghesi, J.-P. Bouchaud, {\it Of songs and men: a Model for Multiple Choice with Herding}, Qual. Quant. 41, 557 (2007)

\bibitem{Shelling} T. Schelling, {\it Micromotives and Macrobehaviour}. W.W. Norton \& Co Ltd (1978)

\bibitem{Buchanan} M. Buchanan, {\it The Social Atom}, Bloomsbury Press, New York, 2007

\bibitem{Ball} P. Ball, {\it Why Society is a Complex Matter}, Springer (2012)

\bibitem{Fortunato} C. Castellano, S. Fortunato, V. Loreto, {\it Statistical physics of social dynamics}, Rev. Mod. Phys, 81, 591 (2009)

\bibitem{Crises} J. P. Bouchaud, {\it Crises and Collective Socio-Economic Phenomena: Simple Models and Challenges}, Journal of Statistical Physics, 151, Issue 3-4, pp 567-606, 2013.

 
\bibitem{spain-new} J. Fernandez-Gracia, K. Suchecki, J. J. Ramasco, M. San Miguel, V. M. Eguíluz, {\it Is the Voter Model a model for voters?}
http://arxiv.org/abs/1309.1131

\bibitem{EW} S. F. Edwards, D. R. Wilkinson, {\it The Surface Statistics of a Granular Aggregate}, Proc. R. Soc. Lond. A  381, 17-31 (1982)

\bibitem{Schweitzer} F. Schweitzer, J. A. Holyst, {\it Modelling Collective Opinion Formation by Means of Active
Brownian Particles}, Eur. Phys. J. B 15, 723 (2000); F. Schweitzer, {\it Coordination of Decisions in a Spatial Model of Brownian Agents}, Economics with Heterogeneous Interacting Agents (WEHIA), Berlin: Springer (2002)

\bibitem{Bourdieu} P. Bourdieu, L. Wacquant, {\it An Invitation to Reflexive Sociology}, University of Chicago Press and Polity, 1992.

\bibitem{Latour} B. Latour, {\it Reassembling the Social. An Introduction to Actor-Network-Theory}, Oxford University Press, 2005.

\bibitem{Weisbuch} G. Deffuant, D. Neau, F. Amblard, and G. Weisbuch, {\it Mixing beliefs among interacting agents}
Adv. Complex Syst. 3, 87-98 (2001)

\bibitem{Anderson} S. P. Anderson, A. De Palma, J. F. Thisse  {\it Discrete Choice Theory of Product Differentiation} (MIT Press, 1992)

\bibitem{spain} M. Pineda, R. Toral and E. Hernandez-Garcia, {\it Diffusing opinions in bounded confidence processes}, Eur.
Phys. J. D 62 109-117 (2011)

\bibitem{spain2} M. Pineda, R. Toral and E. Hernandez-Garcia, {\it Noisy continuous-opinion dynamics}, J. Stat.
Mech. P08001 (2009)

\bibitem{Chakraborti} M. Lallouache, A. S. Chakrabarti, A. Chakraborti and
B. K. Chakrabarti. {\it Opinion formation in kinetic exchange models: Spontaneous
symmetry-breaking transition}, Physical Review E, 82, 056112, 2010


\bibitem{Boltzmann} see e.g. S. Chapman, T. G. Cowling, {\it The mathematical theory of non-uniform gases}
3rd edition. Cambridge University Press, Cambridge (1971).

\bibitem{Grauwin} S. Grauwin, P. Jensen, {\it Opinion groups formation and dynamics: structures that last from non lasting entities}, Phys. Rev. E 85, 066113 (2012)

\bibitem{Cox} K. R. Cox, {\it The Voting Decision in a Spatial Context}, in C. Board
et al., eds, Progress in Geography, Volume 1 (London: Edward Arnold, 1969); P. J. Taylor and P. J. Johnston,
{\it Geography of Elections} (London: Penguin, 1979)

\bibitem{McAlister} I. Mac Allister, R. J. Johston, C. J. Pattie, H. Tunstall, D. F. L. Dorling, D. J. Rossiter,  {\it  Class Dealignment and the Neighbourhood
Effect: Miller revisited},
British Journal of Political Science, {\bf 31}, 41-59 (2001)

\bibitem{andersen} Robert Andersen and Anthony Heath, {\it Social Class and Voting:
A Multi-Level Analysis of Individual
And Constituency Differences}, Centre for research into elections and social trends, Working Paper 83, September 2000

\bibitem{jargowsky} Paul A. Jargowsky, {\it Take the money and run: Economic segregation in US metropolitan areas}, American Sociological Review {\bf 61} 984 (1996)

\end{thebibliography}
\end{document}